\documentclass[epj]{svjour}
\usepackage{graphics}
\begin{document}
\title{Two-dimensional discrete wavelet analysis of \\
multiparticle event topology in heavy ion collisions}

\author{I.~M.~Dremin \inst{1},
G.~Kh.~Eyyubova\inst{2},
V.~L.~Korotkikh\inst{2}
 and L.~I.~Sarycheva\inst{2}
}                     
%
%
\institute{Lebedev Physical Institute, 119991 Moscow, Russia \and Scobeltsyn Institute of Nuclear Physics,
Moscow State University, 119992 Moscow, Russia}

\date{Received: date / Revised version: date}
%
\abstract{
The event-by-event analysis of multiparticle production in high energy hadron 
and nuclei collisions can be performed using the discrete wavelet transformation. 
The ring-like and jet-like structures in two-dimensional angular histograms
 are well extracted  by wavelet analysis. For the first time the method is 
applied to the jet-like events with background simulated by event generators, 
which are developed to describe nucleus-nucleus collisions at LHC energies.
The jet positions are located quite well by the discrete wavelet transformation
of angular particle distribution even in presence of strong background.
} 
\authorrunning{Dremin}
\titlerunning{wavelet analysis of event topology}
\maketitle
\section*{Introduction}
\label{intro}
Physics of high energy nucleus-nucleus collisions is much more complex than a 
mere independent superposition of nucleon-nucleon collisions. 
Already now the first results from RHIC on AuAu collisions at $\sqrt
s = 200$A  GeV  evidenced  collective effects.
The detailed analysis of RHIC results is presented in review papers of all 
four experiments BRAHMS \cite{BRAHMS},
PHOBOS \cite{PHOBOS}, STAR \cite{STAR}, PHENIX \cite{PHENIX}.
 It is concluded 
 that the matter, which is formed, is not described in terms of a color-neutral 
system of hadrons. This state of matter  undergoes the most dense stage reminding a liquid. It  is called quark-gluon liquid or strong interacting QGP (sQGP). Manifestation of this new matter can lead to such unexpected effects as multiple gluon minijet production, asymmetric and odd number of quark jets, collective effects.   
 
The most spectacular structure in the angular distribution of created 
particles are jets of hadrons well collimated along the jet axis.
 Comparison of jet production in AA and pp collisions or in the central and 
peripheral nucleus-nucleus collisions allows
  to estimate jet absorption in the partonic matter. This is one of the main 
signatures of QGP \cite{vard}.

Among suggested QGP signatures, the event-by-event topology plays a special 
role because fluctuations and correlations in angular particle distribution 
provide important information about the dynamics of a process.  
An example of such kind of structure is two particle angular correlations in 
the phase space at low transverse momenta $p_T< 2$ GeV/c which shows that 
jet-like structure  exists even in soft hadron correlations and depends 
on collision centrality \cite{STAR}. The collective azimuthal flow is seen in
individual events as their elliptic ("cucumber-like") shape. 
The ring-like structure of events has been observed \cite{apan,dremin,ajit}. 
This phenomenon is interpreted as Cherenkov gluons \cite{d1,drem1,drem2}
or Mach waves \cite{sto,st,cas} induced by a parton in a medium.

The emission of mini-jets can be another origin of inhomogeneities in particle 
distribution. The clusters of particles with relative small transverse 
energy are interpreted  as  mini-jets \cite {low_et}. 
Mini-jet physics is the new trend that allows to investigate hadron 
fluctuations at low transverse momenta $p_T $. At high $p_T$ 
the jet production holds memory about early hard partonic scatterings 
and jet modifications due to the medium. Mini-jets at lower $p_T$ are expected 
to have shorter mean free path in the medium and thus are more likely to 
dissipate, erasing initial correlations and changing substantionally their 
characteristics in a medium. Comparison of mini-jet event topology 
  in central and peripheral nucleus-nucleus collisions allows to define more 
precisely the density and size of partonic medium and  
 control a degree of its thermodynamic equilibrium \cite {Adams}.

The analysis of event structure by the  wavelet transformation is 
very fruitful for such processes. The wavelet transformation was applied 
both to one- \cite{georg,motal} and two-dimensional \cite{ast,dremin,uzhinskii} particle distributions in nucleus-nuc\-leus collisions.
The discrete wavelet transformation (DWT) \cite {dremin} and continuous 
wavelet transformation \cite {ast,uzhinskii} were used. 
It was shown that the wavelet analysis allows to reveal fluctuation patterns.  
The "texture" of some events in AA collisions was investigated with the help 
of DWT  \cite {berden,Kopytin,Adams}. 

In the future experiments at LHC energies the jet topology of events can give 
the information about characteristics of sQGP, produced in nucleus-nuc\-leus 
collisions. It is necessary to reconstruct energy and position  of jets with 
good resolution. Jet finding algorithms of cone type select
groups of particles within a cone of the certain radius 
$R_{jet}=\sqrt{(\bigtriangleup \eta)^2+(\bigtriangleup \varphi)^2}$ in 
pseudorapidity-azimuthal angle ($\eta \times \varphi $) space \cite {arnison}.
This algorithm can't extract more complex structures in two-dimensional 
angular distributions like rings. We show that this can be done by the wavelet 
analysis of a single event. 

Jet reconstruction by usual methods becomes difficult in central PbPb collisions 
at LHC energy \cite {lokhtin} because of large  background with
$(dN_{ch}/dy)_{y=0} =3000\div 5000 $ charged particles.
Therefore usual cone type algorithms are modified in different ways to take 
into account this background. In \cite {Pb_Pb} the possibility of jet 
recognition in central PbPb collision in CMS detector with the help of 
modified UA1-algorithm was studied. It is shown that for jets with $E_T> 50 $ 
GeV the recognition efficiency is close to 100 \%. 
   
The two dimensional angular distribution in this work is simulated for central 
PbPb collisions at LHC energy by PYTHIA  \cite{pythia} and HYDJET event 
generators \cite {lokhtin}, which are aimed on the comparison with future
LHC data on nucleus-nucleus collisions.

With this goal in mind we try to find if the wavelet transformation helps 
to answer the following questions: 

1. Is it possible to distinguish different structures in the same event 
analysing the shape of their angular distributions? 

2. How to find and separate the hadron jets with different widths?  

3. What is a criterion of jet recognition over a huge background in heavy ion 
multiparticle event? 

 We show that the method of the wavelet analysis allows to distinguish jets
from rings and to find the jet 
positions among a huge background even at rather low jet energies.


\section*{The discrete wavelet transformation}
\label{DWT}
In many papers on wavelets \cite {Malla,ten_lect,drem0} it is shown that for 
expansion of a two-dimensional function $f (x, y) $ in a wavelet basis 
one can use the one-dimensional expansion coefficients. Therefore here we 
explain briefly the basic concepts for expansion and 
restoration of the one-dimensional function $f(x)$.

There is the complete orthonormal basis of scaling $\{\varphi_{j,n}(x)\}$ and 
wavelet $\{\psi_{j,n}(x)\}$ functions with compact support in which one can 
expand any measurable function as a series: 
\begin{equation}
\label{f.1}
f(x) = \sum\limits_{n=-\infty}^{\infty} a_{j_{0}}[n]  \varphi_{j_{0},n}(x) +
\sum\limits_{j=j_{0}} \sum\limits_{n=-\infty}^{\infty} d_j[n] \psi_{j,n}(x). 
\end{equation}
One may  choose wavelets which are defined by finite number of real coefficients
$h [n] $, called a filter. Coefficients $h [n] $ correspond to the scaling function $ \varphi (x) $, and coefficients $g [n] = (-1) ^ {1-n} h [1-n] $
 to the wavelet function $ \psi (x) $. Then the orthonormal basis is constructed as a set of functions:
\begin{equation}
\label{f.2}
\{\psi_{j,n}(x)\}=2^{j/2}\psi(2^jx-n)\}_{j=0,1, ...}
\end{equation}

Wavelet function $\psi_{j, n}(x)$ expands  $2^j$ times with the scale changing 
and shifts along $x$ axis by $n$.
This allows to find features of $f(x)$ not only on the whole $x$-axis 
(as in Fourier transformation), but also in  narrow regions 
corresponding to smaller scales. The restoration is complete due to completeness 
and orthonormality of the wavelet basis. Further we follow C. Mallat \cite{Malla}.

A great advantage of the discrete wavelet analysis is the opportunity 
to proceed with the so called fast wavelet transformation (FWT) using 
the coefficients $h [n] $ and $g [n] $. Computing is done by the iterative 
procedure and is therefore fast.

Values of the function $f (x)$ on the smallest scale are linked to the coefficients $a_{j, n} \equiv a_j [n]$, and further algorithm of transition to a larger scale is \cite{Malla}:
\begin{equation}
\label{f.3}
\begin{array}{rcl}
a_{j+1}[p] & = & \sum\limits_{n=-\infty}^{\infty}h[n-2p]a_j[n],  \\
d_{j+1}[p] & =& \sum\limits_{n=-\infty}^{\infty}g[n-2p]a_j[n],
\end{array}
\end{equation}
while the  algorithm of restoration is:
\begin{equation}
\label{f.4}
a_j[p]=\sum\limits_{n=-\infty}^{\infty}h[p-2n]a_{j+1}[n]
+\sum\limits_{n=-\infty}^{\infty}g[p-2n]d_{j+1}[n].
\end{equation}
The sums in (\ref {f.3}) and (\ref {f.4}) are finite so the finite number of coefficients $h [n] $ and $g [n]$ is used. 

\begin{figure*}
  \resizebox{1.0\textwidth}{!}{%
  \includegraphics{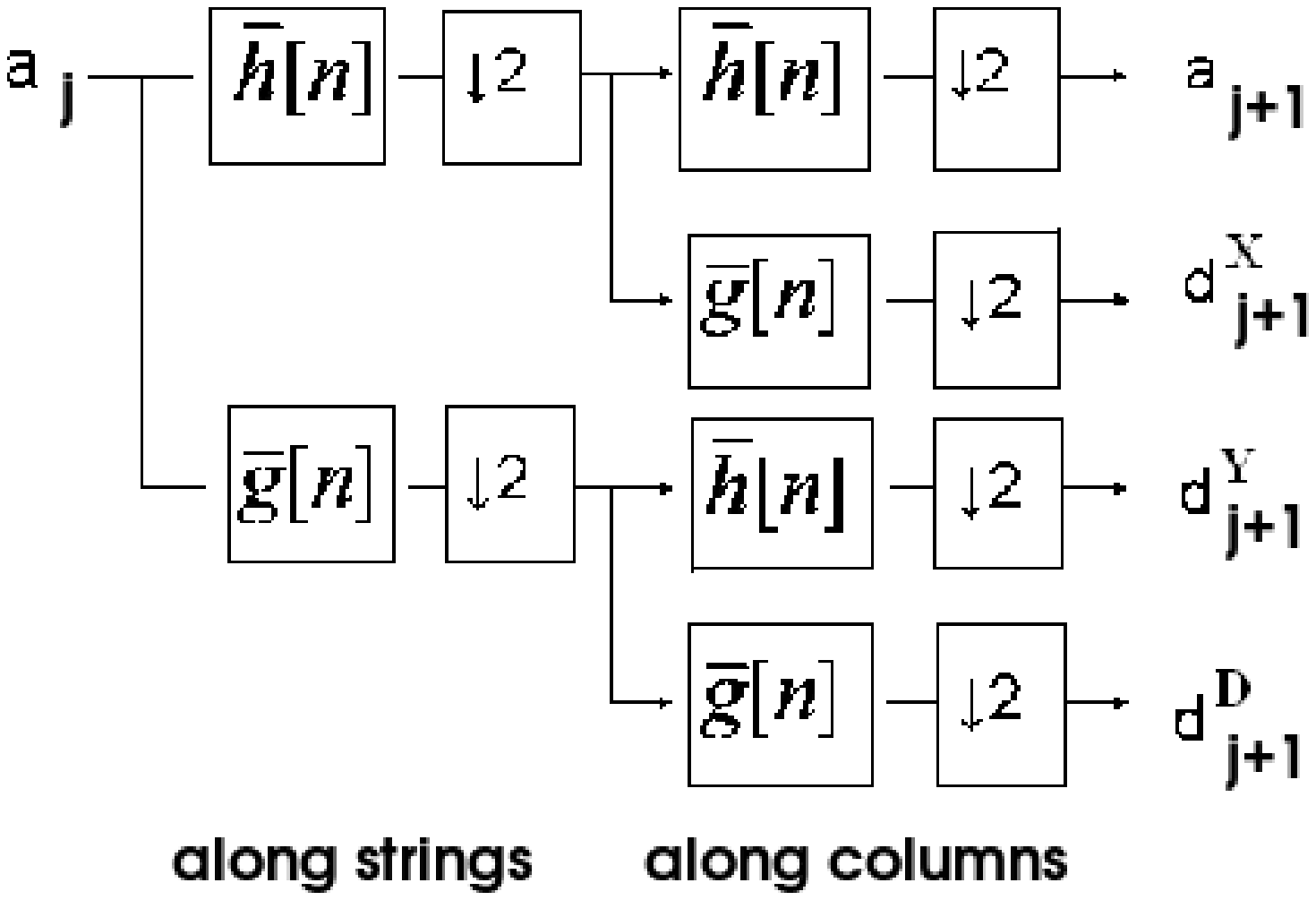}
  \includegraphics{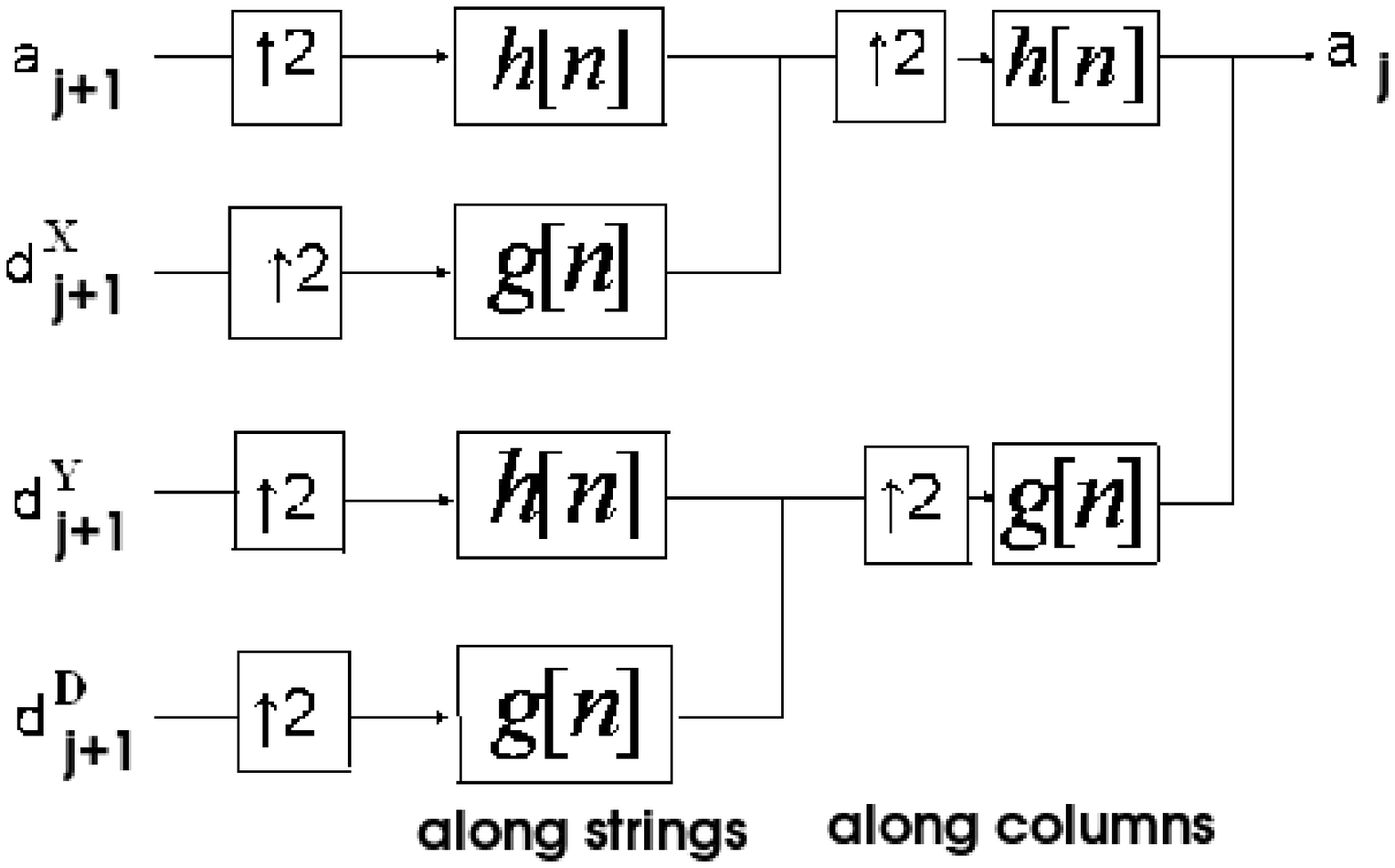}
}
\vspace*{0.4cm}       
 \hskip 50mm a) \hskip 70mm b)
  \caption{Symbolic schemes of the two-dimensional DWT: a) decomposition and 
b) restoration}
\label{fig.1}       
\end{figure*}

In the two-dimensional case a separable form of wavelet functions is used 
\cite{Malla}. Then orthonormal basis is constructed as a set of functions 
$\{\Gamma^X_{j,n_1,n_2},\Gamma^Y_{j,n_1,n_2},\Gamma^D_{j,n_1,n_2}\}$:
\begin{equation}
\label{f.7}
\begin{array}{rcl}
\Gamma^X_{j,n_1,n_2}(x_1,x_2) & = & 2^j\varphi(2^jx_1-n_1)\psi(2^jx_2-n_2); 
\nonumber \\
\Gamma^Y_{j,n_1,n_2}(x_1,x_2) & = & 2^j\psi(2^jx_1-n_1)\varphi(2^jx_2-n_2); 
\nonumber \\
\Gamma^D_{j,n_1,n_2}(x_1,x_2) & = & 2^j\psi(2^jx_1-n_1)\psi(2^jx_2-n_2).
\end{array}
\end{equation}
 Separable two-dimensional convolution can be defined as a product of
 one-dimensional convolutions on the rows and columns.
  Symbolically the decomposition and restoration of a two-dimensional 
function $f(x, y)$ is shown in fig. \ref {fig.1} (see \cite {Malla}).
 Here $\bar h[n]= h[-n]$. Decomposition (fig. \ref {fig.1}a) is done  first 
along strings of table $a_j(x_1,x_2)$, then along columns. 
Restoration (fig. \ref {fig.1}b) is done in the opposite direction.

%

\section*{Wavelet analysis of complex structures in two-dimensional distributions}
\label{WLA}
The wavelet analysis is often called a mathematical microscope since it allows 
to examine a signal with different resolution and to reveal structures on 
different scales. We demonstrate it here with some examples using Daubechies 
wavelet $D^8$ \cite{ten_lect}.  DWT with Daubechies $D^8$ was tested by the 
example of the one-dimensional function  in  \cite{vlk}. 

In figure (fig.\ref {fig.2}a) the two-dimensional histogram of angular 
distribution in $\eta \times \varphi$ space is simulated from $N=100000 $ 
entries, including the jet and ring-like shape as a toy example. 
The histogram binning is $128 \times 128 $. 
The values in each bin of the histogram determine the coefficients 
$a_{j=0}[n_1, n_2]$ for the smallest scale $j=0$. The whole interval of the 
histogram along both axes $\eta$ and $\varphi$  corresponds to the scale $j=7$. 

With the help of the wavelet analysis we try to reveal different forms of 
irregularities in the histogram. For this purpose we use the wavelet 
decomposition of the histogram and then restore its structure at different 
scales. If during restoration we set to zero the coefficients 
$d_{j}[n_1, n_2]=0$ for large scales $j=3,4,5,6,7$, then only a narrow peak 
is restored (fig. \ref {fig.2}b).
If during restoration we set to zero the coefficients $d_{j}[n_1, n_2]=0$ for 
small scales $j=1,2,3$ then
the ring-like structure, which initially is present in the distribution, is 
revealed and the narrow peak disappears (fig. \ref {fig.2}c).

Now we consider another example of the function containing two peaks of 
different shapes:
\begin{eqnarray}
\label{f.9}
f(\eta,\varphi) & = & \frac{1}{\sqrt{(\eta-\eta_0)^2+(\varphi-\varphi_0)^2}}+ \nonumber\\
& +& A\exp\{-0.5\frac{(\eta-\eta_1)^2+(\varphi-\varphi_1)^2}{\sigma^2}\} .
\end{eqnarray}

The chosen parameters are $~\sigma=0.4$ and $A=40$ (the particular values for 
peak positions are irrelevant and chosen for peaks not to overlap).
The peaks have different widths. Therefore they correspond to different scales 
in wavelet expansion. Just as in the previous example, we produce the histogram 
from this function (fig. \ref {fig.3}a). If we leave only coefficients 
$d_{j}[n_1, n_2]$ for $j=1,2$ at small scales, we  get the narrow peak 
(fig. \ref {fig.3}b). The wide peak is restored quite well if in the wavelet 
series we leave coefficients $d_{j}[n_1, n_2]$ at large scales $j=4,5,6$  
(fig.\ref{fig.3}c).

\section*{Wavelet analysis of jet events}
\label{JET}
Jets of hadrons are usually well collimated along the jet axis, so we search by 
wavelet method for narrow peaks in the distribution of the transverse momentum
$\frac{d^2p_T}{d\varphi d\eta}$.
 Jets with different transverse momenta $p_T $ were simulated  for pp 
collisions at LHC energy $ \sqrt {s} = 5500$ GeV by PYTHIA event generator 
\cite {pythia}. 
Then the angular distribution of particle transverse momentum  was superimposed 
on the background, simulated by HYDJET generator \cite {lokhtin}. The 
background is a sum of 100 events of PbPb collisions with multiplicity in the 
central rapidity region $(dN/dy)_{y=0} \approx 4000$.  The jet production is 
switched off in the background events. 

Histograms are made by binning of the two-dimensional distribution within 
intervals $|\eta| < 2.4$ and $|\varphi| < \pi$ with the number of bins 128 for 
each variable. 
 The sizes of bins are equal to $\Delta \eta \times \Delta \varphi= 0.0375 \times 0.049 $ , which are close to CMS detector granularities. 
The wavelet analysis is used to find the position of jets. 
The background is treated as the large-scale structure (in our case $j=2-7 $).

 The following algorithm of jet recognition  with the help of wavelet analysis 
is proposed:

\hskip 5mm 1. A given event distribution is decomposed by the wavelet discrete 
transformation. 

\hskip 5mm 2. A smooth background at the large scale $j=7 $ is evaluated and 
then  subtracted from the distribution.
 
\hskip 5mm 3. To remove the irregularities at the large scales the coefficients 
$d_{j}[n_1,n_2], ~j=4-7$ are set to zero. The range of small scales $j < 4 $ 
corresponds approximately to the region  $\sqrt{(\bigtriangleup \eta)^2+
(\bigtriangleup \varphi)^2} < R_{jet} =0.3$ from the center of the narrow peak.
  
\hskip 5mm 4. The coefficients at small scales $d_{j}[n_1,n_2], j=1,2,3$, which 
are below a certain threshold, are set to zero
in order to remove the sharp irregularities with small intensity. These cuts 
correspond to conditions $d_{j}[n_1,n_2] < cut \times (d_{j}[n_1,n_2])_{max}$. 
The parameter $cut=0.95$ was found as optimum. 

\hskip 5mm 5. The  event distribution is restored  with the selected $d_{j}[n_1,n_2]$  coefficients. 

The event with three jets with total  $p_T=53$ GeV, $p_T=45$ GeV and $p_T=22$ 
GeV in their regions  is shown in fig.\ref {fig.4} before background was added. 
The sum with background is shown in fig.\ref {fig.4b} .  The same event with 
jet reconstruction by the proposed method is shown in fig. \ref {fig.5}.  
 Here the smooth background has been removed by setting to zero 
$d_{j}[n_1,n_2]$  at  largest scale $j=7$. 
The sharp peaks with small intensity in the background are removed by the cuts 
of small scale $d_{j}[n_1,n_2]$-coefficients below the threshold.
Position of jets is well allocated on position of peaks in the spectrum of 
wavelet coefficients at different scales depending on the width of a jet. Fig.
  \ref{fig.5} shows that the peak intensities are decreased and their shapes 
are distorted.

 Only two jets are restored in this case. The third jet with $p_T=22$ GeV/c is 
lost because it is compatible to background fluctuations.
Another disadvantage of this selection is  very strong distortion of the total 
transverse momentum values. 
The calculation of jet energy must be made by  other methods (for example, by 
the cone method). But the jet positions are reconstructed quite well by DWT.

The background under jet may be different and depend on jet  position in $\eta - \varphi$ plane. So it is not correct to subtract background, calculated as average background in   $\eta-\varphi$  space.  
Here we propose to estimate the background contribution under jet by  calculating it
 inside a ring around the jet. Let we know a position of jet by wavelet analysis and select some region of jet as $\sqrt{(\bigtriangleup \eta)^2+(\bigtriangleup \varphi)^2} < R_{jet}$ . Then we can calculate the total transverse momentum  in the ring $R_{jet} < \sqrt{(\bigtriangleup \eta)^2+(\bigtriangleup \varphi)^2} < R_{ring}$.
In our case we take $R_{ring}$ so that the number of bins under  jet and in the ring are equal.


Let's define the average transverse momentum $<p_T> = \frac{1}{N{bin}}~\sum_{k=1}^{N{bin}}p_{T,k}$, 
where $N_{bin}$ is the number of bins in the region of jet or ring in angular 
distribution.
 The first (53 GeV/c) and second (45 GeV/c) jets have the average $<p_T>$  
equal to $<p_T>_{jet}=0.346$ GeV/c and $0.295$ GeV/c. They are larger then those 
for background $<p_T>_{back}=0.330$ GeV/c and $0.246$ GeV/c.
For the third jet (22 GeV/c) in Fig.\ref{fig.5} the average $<p_T>_{jet}=0.143$ 
GeV/c is less then for background ($<p_T>_{back}=0.209$ GeV/c). Thus the 
wavelet analysis can't find this jet.  


The best way to study jet events with large background by wavelet method is as 
follows. First, one finds positions of jets by suggested procedure. Then  the 
background contribution is calculated in the ring region  around each jet. The 
genuine energy of the jet is obtained by the subtraction of the background 
contribution from the  total energy in the jet region. 

\begin{figure}
\begin{center}
\resizebox{0.45\textwidth}{!}{%
  \includegraphics{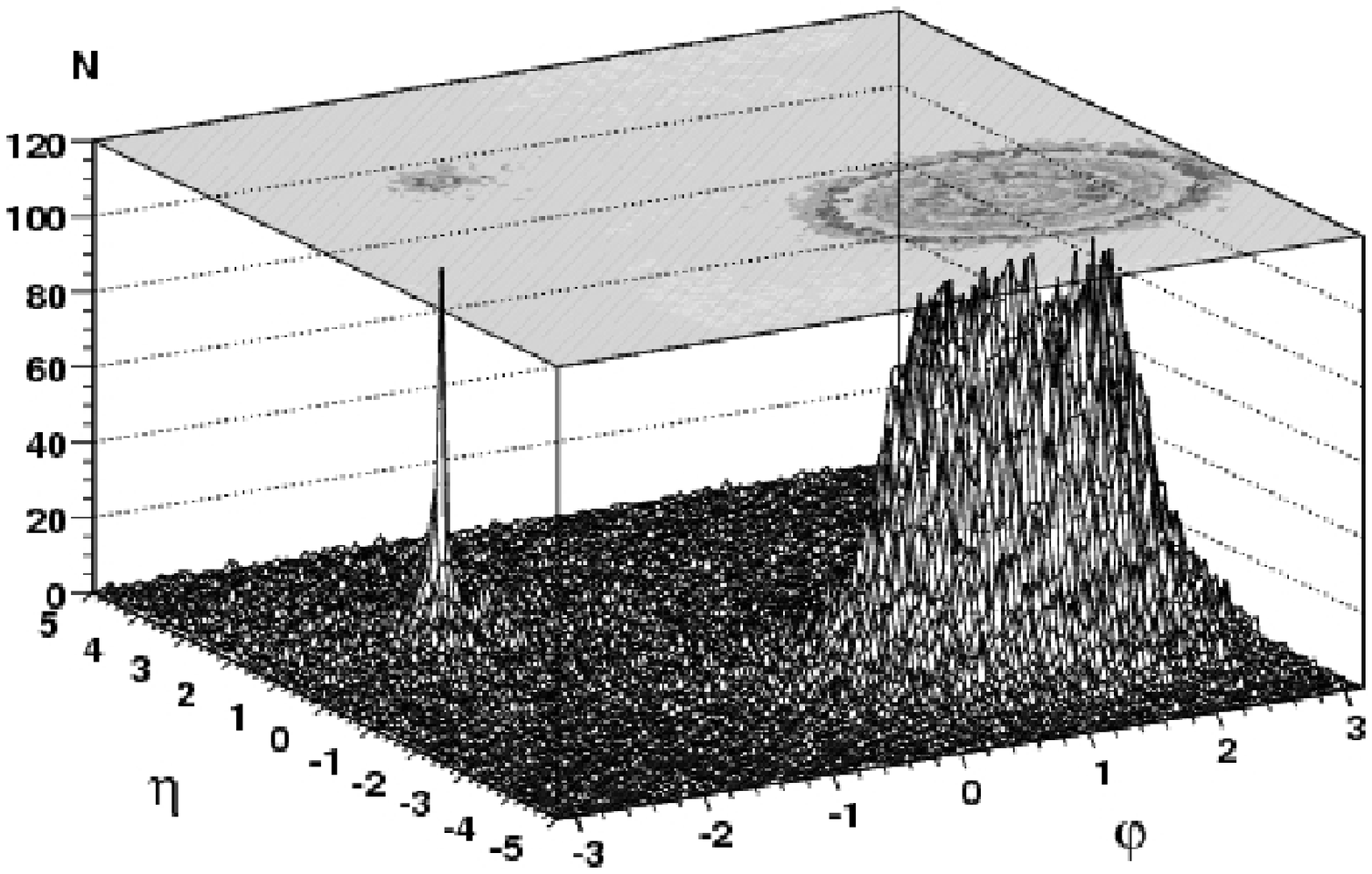}
}
\end{center}
\vspace*{-0.2cm}       
 \centerline{a)} 
\end{figure}

\begin{figure}
\begin{center}
\resizebox{0.43\textwidth}{!}{%
  \includegraphics{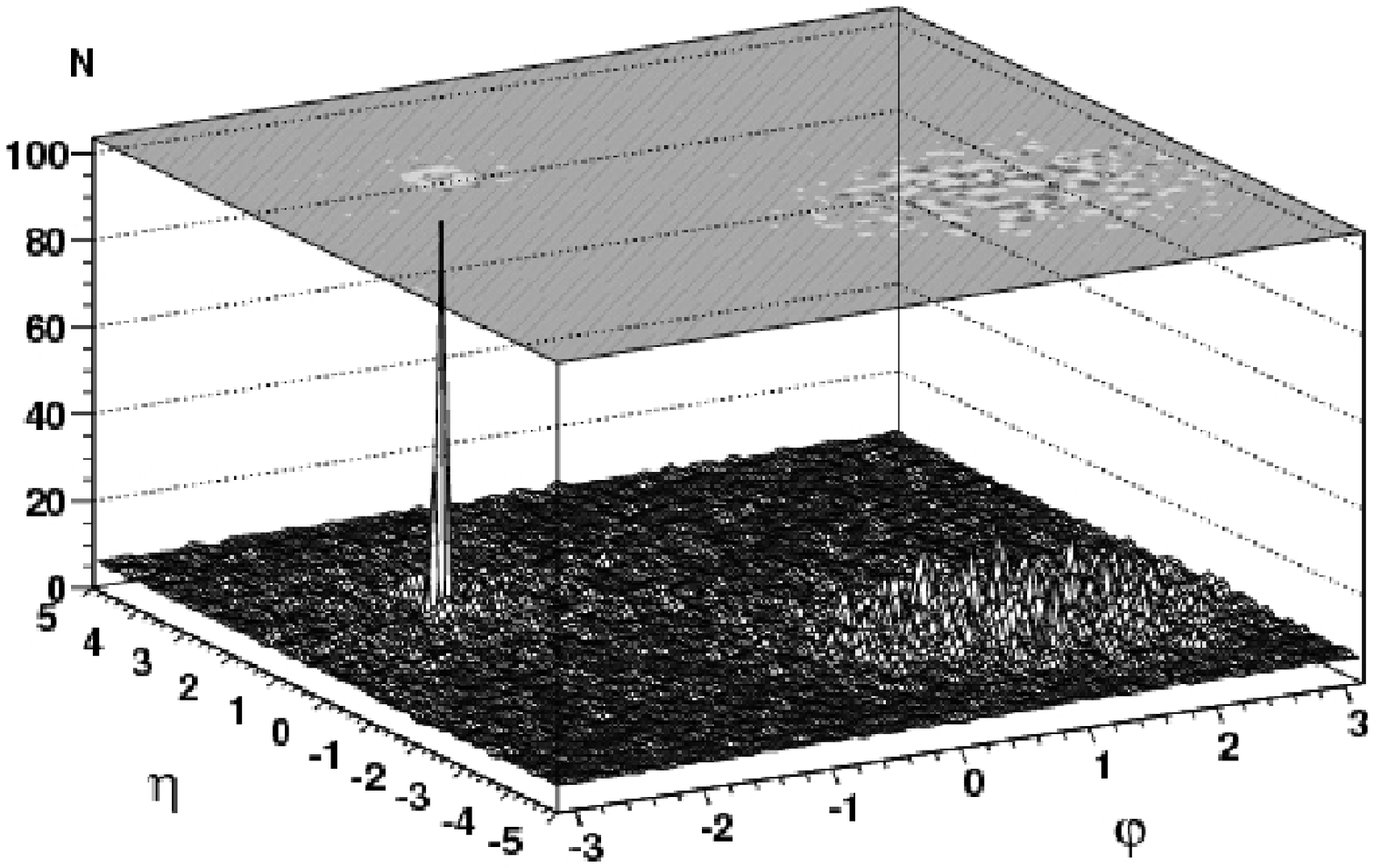}
}
\end{center}
\vspace*{-0.2cm}       
 \centerline{b)}
\end{figure}

\begin{figure}
\begin{center}
\resizebox{0.4\textwidth}{!}{
  \includegraphics{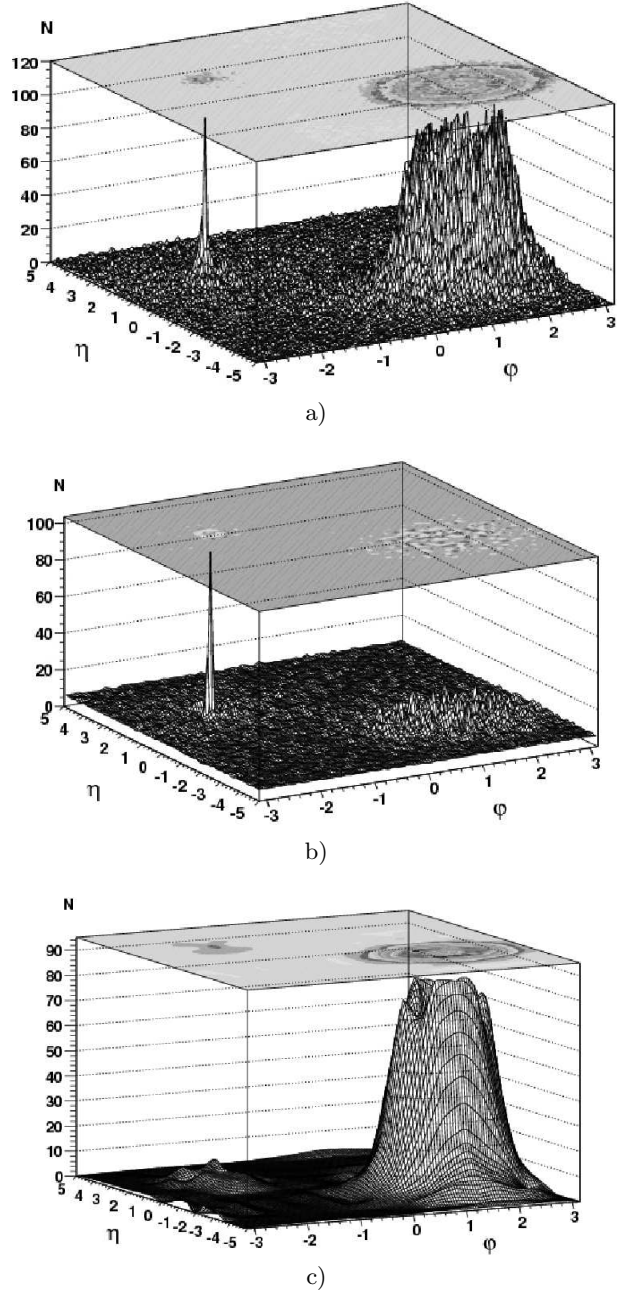}
}
\vspace*{0.3cm}       
 \centerline{c)}
  \caption{Ring-like and jet-like structures in angular particle distribution: 
a) both structures, b) wavelet restoration without large scale coefficients, 
$d_{j}[n_1,n_2]=0$ at $j=3,...,7$,
  c) wavelet restoration without small scale coefficients, $d_{j}[n_1,n_2]=0$ 
at $j=1,...,3$. View in the upper plane of figures is the distribution projected
on the plane.}
\label{fig.2}       
\end{center}
\end{figure}

\section*{Conclusions}
\label{CONCL}
Wavelet analysis allows to find jets and  ring-like structures
in individual high multiplicity events in nucleus-nucleus collisions. 
The two-dimensional discrete wavelet transformation reveals well the 
irregularities with different shapes by corresponding choice of the wavelet 
scale $j$. Such structures  correspond to different effects in nucleus-nucleus 
collisions. The ring-like structure is impossible to reveal by well known jet 
algorithms, but this goal is achieved with the help of the wavelet analysis. 
The method is applied to the jet-like event with high multiplicity background 
for the first time. The DWT allows also to distinguish the bumps  with 
different widths in two-dimensional angular distribution in the same event. 
It is very important for diagnostics of quark and qluon jets.

\begin{figure}
\begin{center}
\resizebox{0.35\textwidth}{!}{%
  \includegraphics{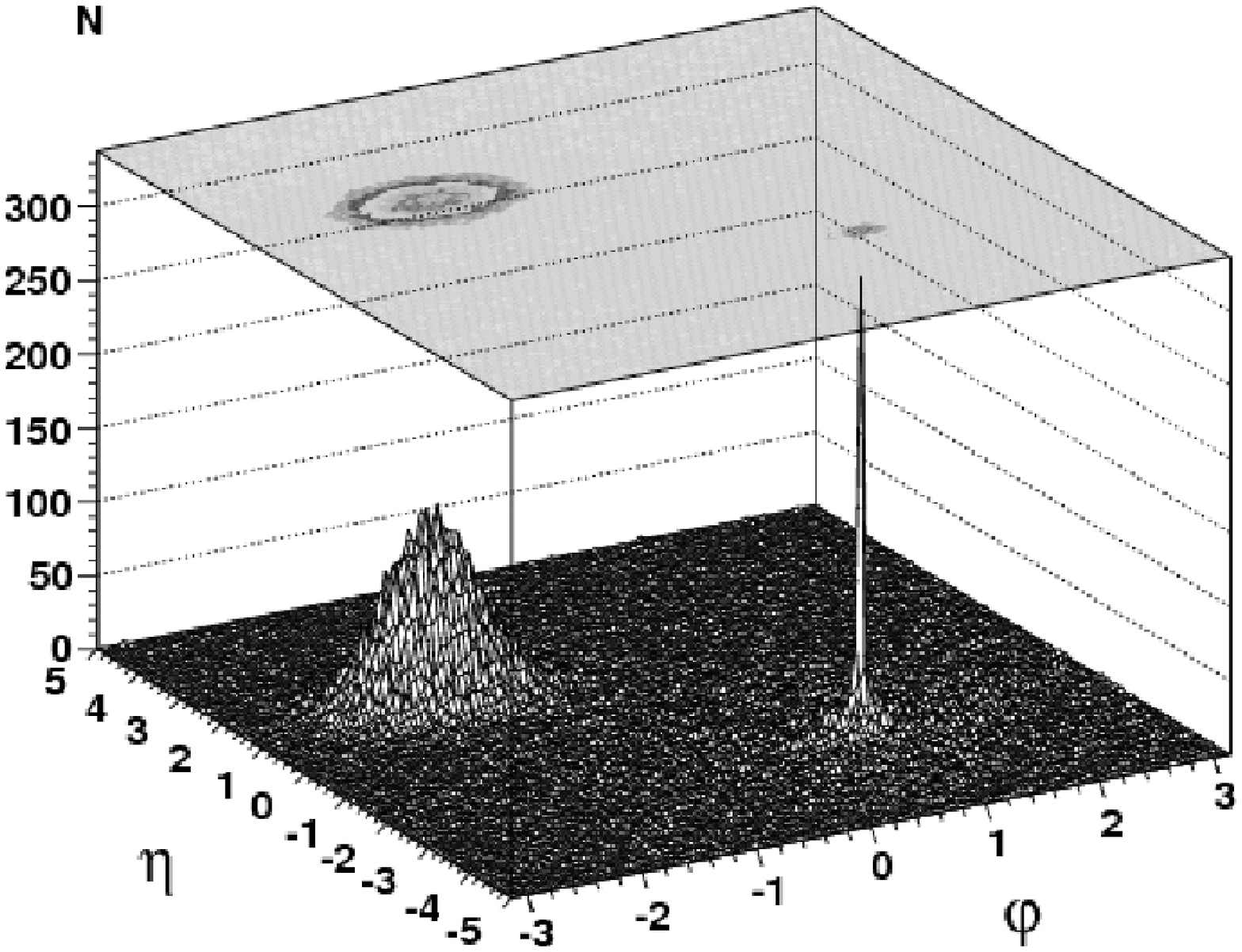}
}
\vspace*{-0.2cm}       
 \centerline{a)} 
\end{center}
\end{figure}
\begin{figure}
\begin{center}
\resizebox{0.35\textwidth}{!}{%
  \includegraphics{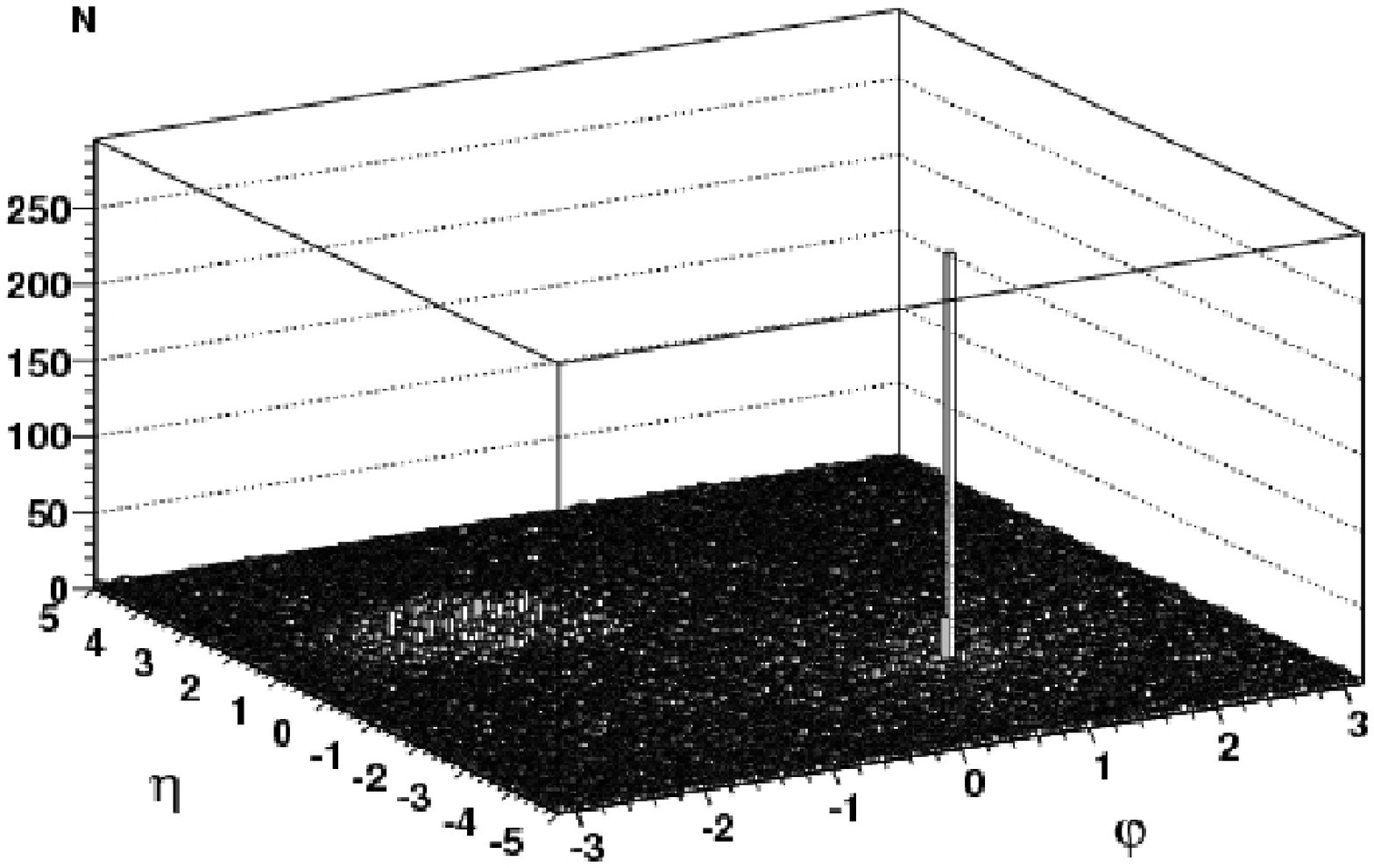}
}
\vspace*{-0.2cm}       
 \centerline{b)}
\end{center}
\end{figure}
\begin{figure}
\begin{center}
\resizebox{0.35\textwidth}{!}{%
  \includegraphics{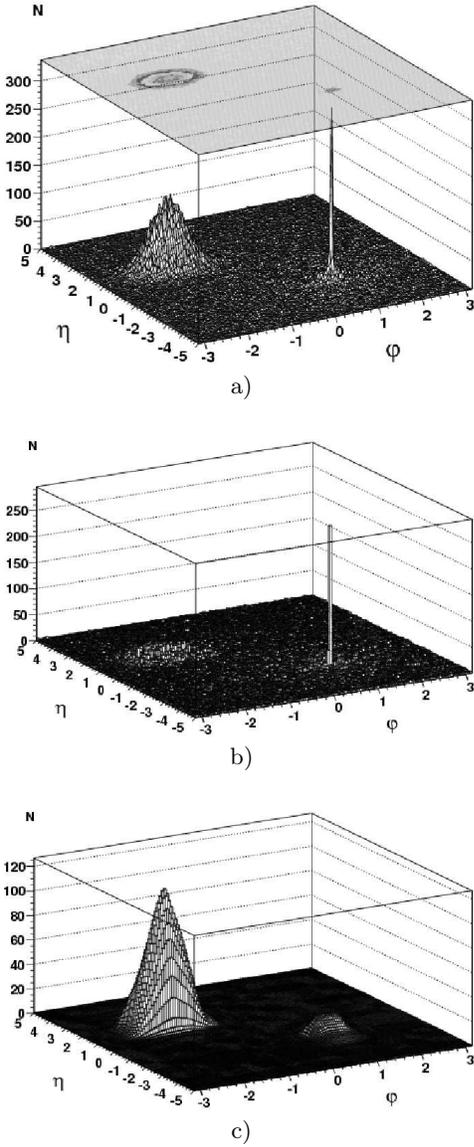}
}
\vspace*{0.2cm}       
 \centerline{c)}
  \caption{Narrow and wide jets in angular particle distribution (\ref{f.9}): 
a) both jets, b) wavelet restoration without large scale coefficients, $d_{j}[n_1,n_2]=0$ at $j=3,...,7$,
c) wavelet restoration without small scale coefficients, $d_{j}[n_1,n_2]=0$ at $j=1,...,3$. }
\label{fig.3}       
\end{center}
\end{figure}

Discrete wavelet transformation was tested on the particle angular 
distributions, which are close to foreseen experimental LHC data.
Position of jets is well allocated on position of peaks in the spectrum of 
wavelet coefficients at different scales depending on the width of a jet.
Removing the  wavelet coefficients at large scales 
allows to substract the smooth background.
Removing the  wavelet coefficients at small scales
below certain threshold allows "to clean out" event from the sharp 
irregularities with small intensity. 

\begin{figure}
\begin{center}
\resizebox{0.35\textwidth}{!}{%
  \includegraphics{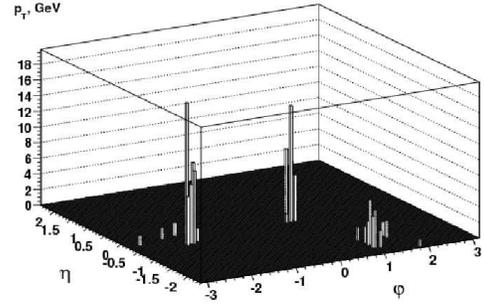}
}
\vspace*{0.4cm}       
 \caption{The event with jet transverse momenta  $p_T=53$ GeV, $p_T=45$ GeV and $p_T=22$ GeV, simulated by PYTHIA.}
\label{fig.4}       
\end{center}
\end{figure}
\begin{figure}
\begin{center}
\resizebox{0.35\textwidth}{!}{%
  \includegraphics{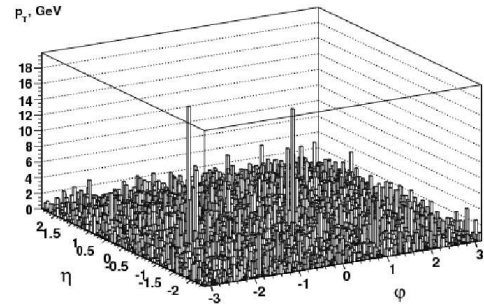}
}
\vspace*{0.4cm}       
  \caption{Sum of the event and the background, calculated by HYDJET. }  
\label{fig.4b}       
\end{center}
\end{figure}
\begin{figure}
\begin{center}
\resizebox{0.35\textwidth}{!}{%
  \includegraphics{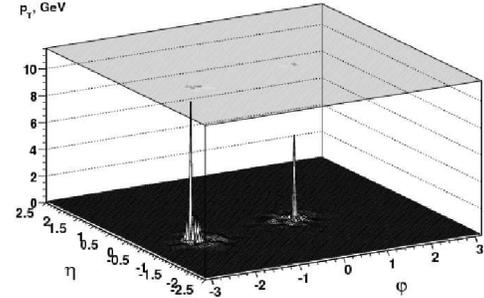}
}
\vspace*{0.4cm}       
  \caption{The distribution of the event (see Fig.\ref{fig.4}) after restoration by suggested method of jet searching.} 
\label{fig.5}       
\end{center}
\end{figure}

One must be careful with excluding the background by wavelet analysis. 
Its subtraction can decrease the jet intensity (energy, number of particles) 
and change the jet shape. It can be recommended for use only to find the jet 
position. The jet energy must be estimated by other methods. However, the jet 
position is defined quite well by the discrete wavelet transformation.

At the end we would like to stress that the wavelet analysis can be used
more widely for studies of the general topology of individual high 
multiplicity events. Beside jets and rings, other structures of mini-bias
events can be found as was noted already. Common features of correlations 
in particle locations within the three-dimensional phase space such as 
fractality and intermittency can be revealed \cite{ddk}.

\section*{Acknowledgements}
\vspace{0.5mm}
We thank  A. M. Snigirev and V. A. Nechitailo  for useful discussions and constructive proposals.


\begin{thebibliography}{}
%
%
\bibitem{BRAHMS} I. Arsene et al. (BRAHMS), Nucl. Phys. {\bf A757} (2005) 1, nucl-ex/0410020.
\bibitem{PHOBOS} B. Back et al. (PHOBOS), Nucl. Phys. {\bf A757} (2005) 28, nucl-ex/050641.
\bibitem{STAR} J. Adams et al. (STAR), Nucl. Phys. {\bf A757} (2005) 102, nucl-ex/0501009.
\bibitem{PHENIX} K. Adcox et al. (PHENIX), Nucl. Phys. {\bf A757} (2005) 184, nucl-ex/0410003 
\bibitem{vard} I.N. Vardanyan et al., Phys. Atom. Nucl. {\bf 68} (2005) 302. 
\bibitem{apan} A.V. Apanasenko et al., JETP Lett. {\bf 30} (1979) 145.
\bibitem{dremin} I.M. Dremin et al., Phys. Lett. {\bf B499} (2001) 97.
\bibitem{ajit} N.N. Ajitanand (PHENIX), nucl-ex/0609038.
\bibitem{d1} I.M. Dremin, JETP Lett. {\bf 30} (1979) 140; Sov. J. Nucl Phys.
{\bf 33} (1981) 726.
\bibitem{drem1} I.M. Dremin, Nucl. Phys. {\bf A767} (2006) 233.
\bibitem{drem2} I.M. Dremin, Int. J. Mod. Phys. {\bf A22} (2007) 3087.
\bibitem{sto} H. St\"ocker et al., Prog. Part. Nucl. Phys. {\bf 4} (1980) 133.
\bibitem{st} H. St\"ocker, Nucl. Phys. {\bf 750} (2005) 123.
\bibitem{cas} J. Casalderrey-Solana et al., Nucl. Phys. {\bf A774} (2006) 577.
\bibitem{low_et} C. Albajar et al., Nucl. Phys. {\bf B309} (1988) 405.
\bibitem{Adams} J. Adams et al., arXiv:nucl-ex/0407001,  arXiv:0706.0596.
\bibitem{georg} G. Georgopoulos et al., Mod. Phys. Lett. {\bf A15} (2000) 1051.
\bibitem{motal}  V. de la Mota1, F. Sa�bille, Acta Physica Hungarica, A) Heavy Ion Physics {\bf 16} (2002) 203. 
\bibitem{ast} N.M. Astafyeva et al., Mod. Phys. Lett. {\bf A12} (1997) 1185.
\bibitem{uzhinskii} V.V. Uzhinskii et al., hep-ex/0206003 (2002).
\bibitem{berden} I. Berden et al., Phys. Rev.  {\bf C65} (2002) 044903.
\bibitem{Kopytin}M. Kopytine, arXiv:nucl-ex/0211015 (2002).
\bibitem{arnison}  G. Arnison et al., (UA1), Phys. Lett.  {\bf B123} (1983) 115.
\bibitem{lokhtin} I.P. Lokhtin, A.M. Snigirev, Eur. Phys. J. {\bf C45} (2006) 211.
\bibitem{Pb_Pb} M. Bedjidian et al., CMS Notes 1996/016. 
\bibitem{pythia} T. Sj\"ostrand et al., Comput. Phys. Commun., 135:238, 2001.
[arXiv:hep-ph/0010017].
\bibitem{Malla} C. Mallat, "A wavelet tour of signal processing", Academic
Press, 1999.
\bibitem{ten_lect} I. Daubeuchies, ``Ten lectures on wavelets'', Soc. for 
Industrial and Applied Mathematics, Philadelphia, Pennsylvania, 1992.
\bibitem{drem0} I.M. Dremin et al., Physics-Uspekhi {\bf 42} (2001) 447, hep-ph/0101182. 
\bibitem{vlk} V.L. Korotkikh,  Preprint  SINP MSU-2002-6/690 (2000).
\bibitem{ddk} E.A. DeWolf et al., Phys. Rep. {\bf 270} (1996) 1.

\end{thebibliography}
\end{document}